\documentclass[prd,draft,onecolumn,showpacs,superscriptaddress]{revtex4}

\usepackage[utf8x]{inputenc}
\usepackage{color}
\usepackage{mathtools}
\usepackage{amsfonts}
\usepackage{hyperref}

\usepackage{graphicx} 

\newcommand{\be}{\begin{eqnarray}}
\newcommand{\ee}{\end{eqnarray}}
\newcommand{\bea}{\begin{eqnarray}}

\newcommand{\eea}{\end{eqnarray}}

\newcommand{\ms}[1]{\textrm{\tiny $#1$}}
\newcommand{\LO}{\ms{(0)}}
\newcommand{\FO}{\ms{(1)}}
\newcommand{\SO}{\ms{(2)}}
\newcommand{\TO}{\ms{(3)}}
\newcommand{\NO}{\ms{(n)}}

\DeclareMathOperator\tr{tr}

\DeclareMathOperator\E{e}
\DeclareMathOperator\Id{Id}

\begin{document}

\title{Canonical charges and asymptotic symmetry algebra of conformal gravity}

\author{Maria Irakleidou}
\email{irakleidou@hep.itp.tuwien.ac.at}
\affiliation{Institute for Theoretical Physics, Technische Universit\"at Wien, Wiedner Hauptstrasse 8--10/136, A-1040 Vienna, Austria}

\author{Iva Lovrekovic}
\email{lovrekovic@hep.itp.tuwien.ac.at}
\affiliation{Institute for Theoretical Physics, Technische Universit\"at Wien, Wiedner Hauptstrasse 8--10/136, A-1040 Vienna, Austria}

\author{Florian Preis}
\email{fpreis@hep.itp.tuwien.ac.at}
\affiliation{Institute for Theoretical Physics, Technische Universit\"at Wien, Wiedner Hauptstrasse 8--10/136, A-1040 Vienna, Austria}

\date{\today}

\preprint{TUW--13--xx}

\begin{abstract} 
We study canonical conformal gravity in four dimensions and construct the gauge generators and the associated charges. Using slightly generalized boundary conditions compared to those in \cite{Grumiller:2013mxa} we find that the charges associated with space-time diffeomorphisms are finite and conserved in time. They are also shown to agree with the Noether charges found in \cite{Grumiller:2013mxa}. However, there exists no charge associated with Weyl transformations. Consequently the asymptotic symmetry algebra is isomorphic to the Lie algebra of the boundary condition preserving diffeomorphisms. For illustrative purposes we apply the results to the Mannheim--Kazanas--Riegert solution of conformal gravity.
\end{abstract}
\maketitle

\section{Introduction} 
Conformal gravity (CG) in four dimensions has recently been studied from various theoretical and phenomenological perspectives. It was shown that it arises as a counter term in five dimensional Einstein gravity (EG) \cite{Liu:1998bu,Balasubramanian:2000pq}
 as well as in twistor string theory \cite{Berkovits:2004jj}. Phenomenologically, it was studied in a series of articles by Mannheim \cite{Mannheim:2012qw,Mannheim:2011ds,Mannheim:2010xw} in  an attempt to describe galactic rotation curves without adding dark matter. Furthermore, CG was used in a cosmological model in \cite{Jizba:2014taa} where the choice of the terms which are responsible for inflation is restricted by demanding conformal invariance.

More generally, conformal invariance is often used as a guiding principle.  For instance, in a series of papers \cite{'tHooft:2009ms,Hooft:2010ac,Hooft:2010nc,Hooft:2014daa} 't Hooft puts emphasis on scale invariance in order to aquire deeper understanding of physics at the Planck scale. Furthermore, recently proposed cosmological models have used Weyl invariant dynamics in scalar-tensor theories as an attempt to explain several cosmological problems. For a recent critical analysis of (fake) conformal symmetry see Ref. \cite{Jackiw:2014koa}.

Historically, CG gained attention after it was shown in \cite{Stelle:1976gc} that higher derivative theories are perturbatively renormalizable. The idea that higher derivative terms might be present in a UV-complete theory of quantum gravity has been first proposed in \cite{Utiyama:1962sn} and put forward in Refs. \cite{'tHooft:1974bx,Deser:1974cz,Deser:1974cy,Deser:1974nb}, where it was shown that the Einstein--Hilbert action is perturbatively non-renormalizable. Furthermore, it was shown in \cite{Goroff:1985th} that Einstein gravity is 2-loop non-renormalizable. Being a higher derivative theory, conformal gravity is one possible candidate one might consider. However,  higher derivative theories in general contain ghosts. For attempts to remove the ghosts in CG see for example \cite{Mannheim:2006rd,Pavsic:2013noa}.
Maldacena has shown in \cite{Maldacena:2011mk} that by using appropriate boundary conditions one can remove the ghosts classically from the theory and at the same time restrict the solutions of CG to those of Einstein gravity.

However, the most general spherically symmetric solution of CG found in \cite{Riegert:1984zz,Mannheim:1988dj} obeys more general boundary conditions. As a step towards a holographic application of CG such weaker boundary conditions were considered in \cite{Grumiller:2013mxa}. These are specified by
\begin{equation}
ds^2=\frac{\ell^2}{\rho^2}\left(d\rho^2+\gamma_{ij}dx^idx^j\right),
\end{equation}
where $\rho\rightarrow 0$ denotes the location of the conformal boundary and $\ell$ is the AdS-radius. Analogous considerations apply to asymptotically dS spacetimes, where Starobinsky boundary conditions \cite{Starobinsky:1982mr} are replaced by a weaker set of consistent boundary conditions. Close to the boundary one assumes a Fefferman--Graham expansion of $\gamma_{ij}$
\begin{equation}
\gamma_{ij}=\sum_{n=0}\gamma_{ij}^{\ms{(n)}}\left(\frac{\rho}{\ell}\right)^n\label{eq:FGexpansion}.
\end{equation}
The variations of the metric are fixed at leading as well as subleading order to be
\begin{eqnarray}
\delta \gamma^{\LO}_{ij}\vert_{\rho\rightarrow 0}=2\lambda \gamma^{\LO}_{ij},\quad \quad \delta \gamma^{\FO}_{ij}\vert_{\rho\rightarrow 0}=\lambda\gamma^{\FO}_{ij},
\end{eqnarray}
with $\lambda$ being regular at the boundary. All higher order terms are allowed to vary freely. The boundary conditions chosen in \cite{Maldacena:2011mk} correspond to setting $\gamma^{\FO}_{ij}=0$. It was shown that with the weaker boundary conditions CG has a well defined variational principle and finite response functions without requiring neither the presence of a Gibbons--Hawking--York like boundary term nor a holographic counter term. Consequently, now there exist two response functions. The first one, $\tau^{ij}$, sourced by $\gamma^{\LO}_{ij}$ corresponds to the Brown--York stress tensor in EG. The second one, $P^{ij}$, is termed partially massless response since it is sourced by $\gamma^{\FO}_{ij}$, which in turn when plugged into the linearized Bach equations \cite{Bach1921} exhibits partial masslessness as defined in \cite{Deser:1983mm,Deser:2001pe}. A recent investigation \cite{Ghodsi:2014hua} of two point correlators showed that in general the theory is non-unitary -- an anticipated feature since CG in general contains ghosts. While non-unitarity is often an undesirable property, there are interesting examples of non-unitary holography \cite{Grumiller:2008qz,Vafa:2014iua} with possible applications to condensed matter systems \cite{Grumiller:2013at}. There could be similar applications of non-unitary CG holography.

In this work we continue the route towards holographic applications of CG. We complement and supplement the study in \cite{Grumiller:2013mxa} by a thorough analysis of the canonically realized charges of CG. Section \ref{section1} serves to fix our notation and to summarize the Hamiltonian formulation of CG carried out in \cite{Kluson:2013hza}. In section \ref{section2} we apply the so called Castellani algorithm in order to obtain the gauge generators on the phase space. After specifying the boundary conditions and studying the boundary condition preserving transformations we turn to the main purpose of this artice in section \ref{section3}. There we derive the charges associated with the gauge generators by requiring that the generators are functionally differentiable. We prove that the charges are finite and conserved in time. Furthermore, we also show that they are equivalent to the charges found in \cite{Grumiller:2013mxa}. In the case at hand we end up with an asymptotic symmetry algebra that is isomorphic to the Lie algebra of the boundary condition preserving transformations (consequently without any central extensions). Finally we summarize our findings and discuss possible directions for future research in section \ref{section4}.

\section{Canonical conformal gravity}\label{section1}
The Lagrangean density of four dimensional conformal gravity is given by
\begin{equation}
\mathcal{L}=-\frac{1}{4}\omega_gC^a_{\ bcd}C_a^{\ bcd}
\end{equation}
with $\omega_g$ being the volume form $\sqrt{\vert g\vert}d^4x$ of the metric tensor $g_{ab}$ on the 4d manifold $\mathcal{M}$ and $C^a_{\ bcd}$ denotes the Weyl tensor, which is traceless and additionally inherits the symmetries of the Riemann tensor. Furthermore, with this particular index position the Weyl tensor is invariant under rescalings of the metric
\begin{equation}
g_{ab}\rightarrow \Omega^2g_{ab},
\end{equation}
Therefore, due to the scaling of the volume form together with that of the metric and of the three inverse metrics that contract the Weyl tensors the Lagrangean is also invariant under these rescalings. Hence CG has an additional gauge symmetry as compared to Einstein's theory of general relativity. As a consequence the theory is only sensitive to Lorentz angles encoded in the metric but not to distances.

\subsection{Arnowitt--Deser--Misner decomposition}
For the Hamiltonian formulation of CG we will follow the study presented in \cite{Kluson:2013hza}. For earlier work on the canonical analysis of higher derivative theories consult Refs. \cite{Kaku:1982xt,Boulware:1983yj,Buchbinder:1987vp}. To that end we perform the standard Arnowitt--Deser--Misner (ADM) decomposition of spacetime \cite{Arnowitt:1962hi} by specifying a function $t$ on $\mathcal{M}$, referred to as the time which by assumption allows to foliate $\mathcal{M}$ with a family of spatial hypersurfaces $\Sigma_t$, defined by $t=\mathrm{const}$. The tangent bundle $\mathcal{T}\Sigma$ can then be identified with the kernel of the 1-form $\nabla_at$, where $\nabla$ denotes the Levi-Cevita connection on $\mathcal{M}$. The hypersurfaces are spatial if $g^{ab}\nabla_at\nabla_bt<0$. The future pointing hypersurface normal is $n_a=\alpha\nabla_a t$ with a normalization constant $\alpha$, specified as follows: Assume there exists a congruence of curves with tanget vector fields $t^a$ such that $t^a\nabla_a t=1$. We decompose this field by 
\begin{equation}
t^a=Nn^a+N^a,
\end{equation}
with $N$ being the so called lapse function, measuring the tick rate of a physical observer following $n^a$, and $N^a$ denoting the shift vector field describing the drag of the coordinate grid orthogonal to $n^a=0$. Then we conclude that $t^an_a=\alpha=-N$, hence $n_a=-N\nabla_at$.

The $3+1$ (ADM) decomposition of the metric reads
\begin{equation}
g_{ab}=-n_an_b+h_{ab}
\end{equation}
with $h_{ab}$ denoting the metric on $\Sigma$. The object $h_a^{\ b}$ can be used as a projector to the tangent bundle $\mathcal{T}\Sigma$ and the cotangent bundle $\mathcal{T}_\ast\Sigma$. The corresponding Levi-Cevita connection will be denoted by $D_a$. Tensor fields $P$ over $\Sigma$ can be obtained from tensor fields $\mathcal{P}$ on $\mathcal{M}$ by using the afore-mentioned projector, which we denote by
\begin{equation}
P_{a_1\cdots a_n}=\perp\!\mathcal{P}_{a_1\cdots a_n}:=h_{a_1}^{\ b_1}\cdots h_{a_n}^{\ b_n}\mathcal{P}_{b_1\cdots b_n}.
\end{equation}
 The action of $D_a$ on intrinsic tensor fields is related to the Levi-Cevita connection on $\mathcal{M}$ by 
\begin{equation}
DP=\perp[\nabla (\perp\mathcal{P})],
\end{equation}
where we suppressed the indices. The square root of the determinant is decomposed as $\sqrt{g}=N\sqrt{h}$, hence $\omega_g=Ndt\wedge \omega_h$.

The extrinsic curvature of the spatial hypersurfaces is given by
\begin{equation}
K_{ab}=h_a^{\ c}\nabla_cn_b= \frac{1}{2}\,^{\ms{(4)}}\pounds_n h_{ab},
\end{equation}
where ${}^{\ms{(4)}}\!\pounds_n$ denotes the Lie derivative along $n^a$ on the manifold $\mathcal{M}$. Lie derivatives on $\Sigma$ will be written without the prefix. The tensor $K_{ab}$ describes the change of $n^a$ projected to the hypersurface, i.e. how the surface bends and curves with respect to the space it is embedded in. Note that since $n_a$ is proportional to a gradient field we have $K_{ab}=K_{ba}$, i.e. following neighboring $n^a$ one measures expansion and shear but no vorticity.

For any covariant tensor field $P_{a_1\cdots a_n}$ on $\Sigma$ we have
\begin{equation}
n^{a_i}\,{}^{\ms{(4)}}\pounds_n P_{a_1\cdots a_i \cdots a_n}=- P_{a_1\cdots a_i \cdots a_n}(n\nabla)n^{a_i}+ P_{a_1\cdots a_i \cdots a_n}(n\nabla)n^{a_i}=0.
\end{equation}
Hence the Lie derivative with respect to $n^a$ on any covariant spatial tensor field is spatial. Also note that for any spatial tensor field $P_{a_1\cdots a_n}$ the four dimensional Lie derivative with respect to a spatial vector field $V^a$ becomes the three dimensional (induced) Lie derivative on $\Sigma$ only \textit{after} the projection to the tensor bundle on $\Sigma$ is carried out
\begin{equation}
\perp\!{}^{\ms{(4)}}\pounds_V P_{a_1\cdots a_i \cdots a_n}=\pounds_V P_{a_1\cdots a_i \cdots a_n}.
\end{equation}
This is vital when defining velocities, i.e. the change of a spatial quantity as $t$ changes \textit{measured on the spatial slice}. For example
\begin{eqnarray}
\dot{h}_{ab}&=&\perp\!{}^{\ms{(4)}}\pounds_t h_{ab}=N\,^{\ms{(4)}}\pounds_n h_{ab}+\pounds_N h_{ab}=\nonumber\\
&=&2\left(NK_{ab}+D_{(a}N_{b)}\right).
\end{eqnarray}

The decomposition of the Weyl tensor is worked out in detail in Appendix \ref{ADMWeyl}. One finds the ADM split of the Lagrangean to be
\begin{equation}
\mathcal{L}=N\omega_h\left(\perp\!n^eC_{ebcd}\perp\!n_fC^{fbcd}-2\perp\!n^en^fC_{aecf}\perp\!n_gn_hC^{agch}\right).
\end{equation}
In contrast to the ADM formulation of general relativity the velocity of $K_{ab}$, i.e. the acceleration of $h_{ab}$ does not drop out of the Lagrangean. Therefore we will regard $K_{ab}$ as an independent canonical coordinate and introduce the relation of $K_{ab}$ and $\dot{h}_{ab}$ as a constraint supplemented with a new auxiliary $\lambda^{ab}$ field.
\begin{eqnarray}
\mathcal{L}=N\omega_h\left\{ -\frac{1}{2}\mathcal{T}^{abcd}\left[R_{ab}+K_{ab}K-\frac{1}{N}\left(\dot{K}_{ab}-\pounds_NK_{ab}-D_aD_bN\right)\right]\times\right.\nonumber\\
\left[R_{cd}+K_{cd}K-\frac{1}{N}\left(\dot{K}_{cd}-\pounds_NK_{cd}-D_cD_dN\right)\right]+\nonumber\\ 
\left.B_{abc}B^{abc}+\lambda^{ab}\left[\frac{1}{N}\left(\dot{h}_{ab}-\pounds_Nh_{ab}\right)-2K_{ab}\right]\right\},\label{Lagrangean1}
\end{eqnarray}
where $\mathcal{T}^{abcd}$ denotes the projector to traceless 2-valent symmetric tensor fields defined in \eqref{tracelessproj} and $B_{abc}$ is the magnetic part of the Weyl tensor defined in \eqref{magneticWeyl} with respect to $n^a$.

In order to circumvent the appearance of second class class constraints and the necessity of introducing Dirac brackets we treat the auxiliary field $\lambda^{ab}$ following the Faddeev--Jackiw method \cite{Faddeev:1988qp}. The form of the Lagrangean in Eq. \eqref{Lagrangean1} allows to immediatly identify the momenta conjugate to $h_{ab}$ and $K_{ab}$ denoted by $\Pi_h^{ab}$ and $\Pi^K_{ab}$ and convert the Lagrangean into the equivalent first order form
\begin{eqnarray}
\mathcal{L}&=&\Pi_K^{ab}\dot{K}_{ab}+\Pi_h^{ab}\dot{h}_{ab}+N\left[\omega_h^{-1}\frac{\Pi_K^{ab}\Pi^K_{ab}}{2}-\Pi_K^{ab}\left(R_{ab}+K_{ab}K\right)+\omega_hB_{abc}B^{abc}-2\Pi_h^{ab}K_{ab}\right]\nonumber\\
&\ &-\Pi_K^{ab}D_aD_bN-\Pi_K^{ab}\pounds_NK_{ab}-\Pi_h^{ab}\pounds_Nh_{ab}-\lambda_P\Pi_K^{ab}h_{ab}.\label{Lagrangean2}
\end{eqnarray}
The last term is necessary to ensure that $\Pi_K^{ab}$ is traceless. Note that by partial integration the Lagrangean can be brought into the following form
\begin{eqnarray}
L&=&\int_\Sigma \left\{\Pi_K^{ab}\dot{K}_{ab}+\Pi_h^{ab}\dot{h}_{ab}-N\left[-\omega_h^{-1}\frac{\Pi_K^{ab}\Pi^K_{ab}}{2}+\Pi_K^{ab}\left(R_{ab}+K_{ab}K\right)-\omega_hB_{abc}B^{abc}+2\Pi_h^{ab}K_{ab}+D_aD_b\Pi_K^{ab}\right]+\right.\nonumber\\
&\ &\quad\left.-N^c\left[\Pi_K^{ab}D_cK_{ab}-2D_a\left(\Pi_K^{ab}K_{bc}\right)-D_a\Pi_h^{ab}h_{bc}\right]-\lambda_P\Pi_K^{ab}h_{ab}\right\}+\nonumber\\
&\ &-\int_{\partial\Sigma}\ast\left[\Pi_K^{ab}D_bN-D_b\Pi_K^{ab}N+2N^c\left(\Pi_h^{ab}h_{bc}+\Pi_K^{ab}K_{bc}\right)\right],\label{Lagrangean3}
\end{eqnarray}
where the $\ast\ $-symbol denotes contraction of the free vector index with the volume form hidden in the momentum variables. In this form one could easily identify the \textit{true constraints} of the Hamiltonian formulation in the sense of \cite{Faddeev:1988qp}. These are the expressions that multiply $N$, $N^a$, and $\lambda_P$, which in turn could be viewed merely as Lagrange multipliers. For later convenience we will only do so in the case of $\lambda_P$ and the primary constraint $\mathcal{P}:=\Pi_K^{ab}h_{ab}$, but not in the case of $N$ and $N^a$ and the associated constraints
\begin{eqnarray}
\mathcal{H}_\perp&=&-\omega_h^{-1}\frac{\Pi_K^{ab}\Pi^K_{ab}}{2}+D_aD_b\Pi_K^{ab}+\Pi_K^{ab}\left(R_{ab}+K_{ab}K\right)-\omega_hB_{abc}B^{abc}+2\Pi_h^{ab}K_{ab},\\
\mathcal{V}_c&=&\Pi_K^{ab}D_cK_{ab}-2D_a\left(\Pi_K^{ab}K_{bc}\right)-D_a\Pi_h^{ab}h_{bc}.
\end{eqnarray}
This means that we keep lapse and shift as canonical coordinates. Hence we find the primary constraints $\Pi_N\approx 0$ and $\Pi^{\vec{N}}_{a}\approx 0$ and consequently identify $\mathcal{H}_\perp$ and $\mathcal{V}_c$ as secondary constraints by demanding that the primary constraints are preserved in time. The advantage of this seemingly superficial step is that the gauge generators found by applying the Castellani algorithm have a proper space--time interpretation, which is useful when we will discuss the asymptotic symmetry algebra of conformal gravity.

\subsection{Hamiltonian formulation}
The total Hamiltonian is the Legendre transform of Eq. \eqref{Lagrangean3} supplemented by the primary constraints and reads
\begin{eqnarray}
H_T=\int_{\Sigma}\left(\lambda_N\Pi_N+\lambda_{\vec{N}}^a\Pi^{\vec{N}}_{a}+\lambda_P\mathcal{P}+N\mathcal{H}_\perp+N^a\mathcal{V}_a\right)+\int_{\partial\Sigma}\left(\mathcal{Q}_\perp+\mathcal{Q}_D\right),
\end{eqnarray}
where the surface contribution is just the negative of the last line in Eq. \eqref{Lagrangean3}. The canonical Poisson brackets are defined by
\begin{equation}
\left\{q_A(x),p^B(x')\right\}=\delta(x-x')\delta^B_A,
\end{equation}
for the respective canonical pairs $\left(h_{ab},\Pi_h^{cd}\right)$, $\left(K_{ab},\Pi_K^{cd}\right)$, $\left(N,\Pi_N\right)$, and $\left(N^a,\Pi^{\vec{N}}_{c}\right)$, where $\delta^B_A$ denotes a symmetrized product of Kronecker deltas.

The consistency of $\mathcal{P}$ also yields one further secondary constraint denoted by $\mathcal{W}$
\begin{equation}
\left\{H_T,\mathcal{P}\right\}=N\left(\Pi_K^{ab}K_{ab}+2\Pi_h^{ab}h_{ab}\right)+NK\mathcal{P}-D_c\left(N^c\mathcal{P}\right)\approx N\left(\Pi_K^{ab}K_{ab}+2\Pi_h^{ab}h_{ab}\right) =:N\mathcal{W}.
\end{equation}
As a next step one can show that the Poisson algebra of the constraints found so far closes (for the variations of the constraints $V$ and $H_\perp$ see Appendix \ref{app}). Thus they are first class in the sense of Dirac. Since the total Hamiltonian consists of constraints only the Dirac algorithm terminates.

The nonvanishing Poisson brackets among the constraints read
\begin{eqnarray}
\left\{V\left[\vec{X}\right],V\left[\vec{Y}\right]\right\}&=&V\left[\pounds_{\vec{X}}\vec{Y}\right],\\
\left\{V\left[\vec{X}\right],H_\perp[\epsilon]\right\}&=&H_\perp\left[\pounds_{\vec{X}}\epsilon\right],\\
\left\{V\left[\vec{X}\right],P[\epsilon]\right\}&=&P\left[\pounds_{\vec{X}}\epsilon\right],\\
\left\{V\left[\vec{X},\right],W[\epsilon]\right\}&=&W\left[\pounds_{\vec{X}}\epsilon\right],\\
\left\{P[\epsilon],W[\eta]\right\}&=&P[\epsilon\eta],\\
\left\{P[\epsilon],H_\perp[\eta]\right\}&=&-W[\epsilon\eta]-P[\epsilon\eta K],\\
\left\{W[\epsilon],H_\perp[\eta]\right\}&=&H_\perp[\epsilon\eta]+P[D^2\epsilon\eta+\epsilon D^2\eta-D\epsilon\cdot D\eta)],\\
\left\{H_\perp[\epsilon],H_\perp[\eta]\right\}&=&V\left[\epsilon D^a\eta-\eta D^a\epsilon\right]+P\left[\left(\epsilon D^a\eta-\eta D^a\epsilon\right)\left(D_cK^c_{\ a}-D_cK\right)\right],\label{eq:DiracAlg}
\end{eqnarray}
where capital roman letters denote the smeared versions of the corresponding constraint densities denoted by calligraphic letters.

Finally one can infer that only 6 physical degrees of freedom remain: we started with $32$ phase space coordinates per point in space and found $10$ first class constraints, eliminating $2\times 10$ phase space coordinates, which leaves $12/2$ physical degrees of freedom. These $6$ degrees of freedom are then usually divided into two contributions: $2$ of these degrees of freedom are provided by a massless and $4$ by a ``partially massless'' graviton.
\section{Gauge generators}\label{section2}
The algorithm developed by Castellani in Ref. \cite{Castellani:1981us} consists of a systematic way to construct all the generators of the gauge symmetries associated with a constraint Hamiltonian theory. A detailed analysis and applications to Yang-Mills theories and general relativity can be found there. Here, we describe briefly this procedure and then we apply it in the case of conformal gravity.

By definition, a gauge symmetry is a symmetry that leaves the action invariant (up to total derivatives) and it is described by transformations of the dynamical variables of the theory. In other words, a gauge symmetry reflects the redundancy in the variables we use to describe a physical state of the theory and therefore this physical state is invariant under the gauge transformations responsible for this redundancy. Such a physical state in a Hamiltonian theory is a trajectory in the phase space that satisfies the equations of motion and the constraints. In order to find the gauge generators, one considers a transformation on the canonical variables and their conjugate momenta that is generated by a phase-space function $G$ and can be parametrized by an arbitrary infinitesimal parameter $\epsilon(t)$. In general, we should also allow for time derivatives of this parameter, $ \epsilon^{\NO}\equiv d^{n} \epsilon / dt$, but of finite order. Of course, this transformation reflects on the trajectory in the phase space giving as a result a varied trajectory. Because we are after the generator of the gauge symmetries of our theory, the varied trajectory must satisfy the equations of motion and the constraints. This gives us a set of conditions for the gauge transformations which can be solved to determine all the gauge generators. In that sense, the Castellani algorithm is exhaustive and provides all the gauge generators. In particular, one uses the ansatz
\begin{equation}
\label{eq:g}
\mathcal{G}(\epsilon, \epsilon^{\FO}, \epsilon^{\SO}, \ldots, \epsilon^{k})= \sum_{n=0}^{k} \epsilon^{\NO} \mathcal{G}_{\NO},
\end{equation}
where $G_{\NO}$ are all the first class constraints of the theory and the last one of the chain $(G_{k})$ is a primary first class constraint (PFC). Because all such chains must end with a PFC the conditions take the form of the following recursive relations
\begin{eqnarray}
\label{eq:ca}
\{\mathcal{G}_{0}, H_{T} \} &=& PFC, \nonumber\\
\mathcal{G}_{0}+ \{\mathcal{G}_{1}, H_{T} \} &=& PFC, \nonumber\\
\vdots \nonumber\\
\mathcal{G}_{k-1} + \{\mathcal{G}_{k}, H_{T}\} &=& PFC, \nonumber\\
\mathcal{G}_{k}&=& PFC,
\end{eqnarray}
where PFC is understood as a linear combination of primary first class constraints. One notices immediately that $k$ gives the number of generations of the secondary constraints, hence in the case at hand $k=1$. In order to find the generators one works the recursive relations presented above in reverse order starting with the PFCs
\begin{equation}
\label{eq:PFC}
\Pi_{N} \approx 0\,\,\,,\,\,{\Pi_{\vec{N}}}_{i} \approx 0\,\,\,,\,\,\frac{\mathcal{P}}{N}\approx 0.
\end{equation}
First we consider $\mathcal{G}_{1}=\Pi_{N}$ and demand
{\setlength\arraycolsep{0.01pt}
\begin{eqnarray}
\label{eq:caone}
\mathcal{G}_{1}&=& \Pi_{N}, \nonumber\\
\mathcal{G}_{0}+\{\Pi_{N}, H_{T} \}&=& PFC, \nonumber\\
 \{\mathcal{G}_{0}, H_{T} \} &=& PFC.
\end{eqnarray}}
The linear combination on the right hand side of the second line in Eq. \eqref{eq:caone} is written as
\begin{equation}
\label{eq:PFCone}
PFC(x)=\int_{\Sigma} \,\,\,\left(\alpha(x, y)\Pi_{N} (y)+\beta^{i}(x,y){\Pi_{\vec{N}}}_{i} (y)+\gamma(x, y)\mathcal{P}(y)\right).
\end{equation}
Solving for the coefficients we find
\begin{eqnarray}
\label{eq:c1}
&&\alpha(x, y)=N^{a}(y)\partial_{a}\delta^{3}(x-y)\,\,\,,\,\,\beta^{a}(x, y)=\delta^{3}(x-y)D^{a}N(y)+N(y)\gamma^{ab}D_{b}\delta^{3}(x-y),\nonumber\\
&&\gamma(x, y)=\frac{\lambda_{\mathcal{P}}}{N}(y)\delta^{3}(x-y),
\end{eqnarray}
from which we construct the canonical gauge generator associated with diffeomorphisms orthogonal to the spatial hypersurface using Eq. \eqref{eq:g}
\begin{equation}
\label{eq:g1}
G_{\perp}[\epsilon, \dot{\epsilon}]= \int_\Sigma \! \left[\dot{\epsilon}^{\FO} \Pi_{N}+ \epsilon \left(\mathcal{H}+\pounds_{\vec{N}}\Pi_{N}+{\Pi_{\vec{N}}}_{a} D^{a}N+D_{a}(\Pi_{\vec{N}}^{a}N)+\frac{\lambda_{\mathcal{P}}}{N}\mathcal{P}\right)\right].
\end{equation}
Similar calculations for $\mathcal{G}_{1a}={\Pi_{\vec{N}}}_{a}$ and $\mathcal{G}_{1}=\mathcal{P}/N$ yield the generators  associated with spatial diffeomorphisms and with Weyl rescalings respectively
\begin{eqnarray}
G_D[\epsilon^{a}, \dot{\epsilon}^{a}]= \int_\Sigma \left[ \dot{\epsilon}^{\FO a} {\Pi_{\vec{N}}}_{a}+ \epsilon^{a} \left(\mathcal{V}_{a}+\Pi_{N} D_{a}N+\pounds_{\vec{N}}{\Pi_{\vec{N}}}_{a}\right)\right],\label{eq:g2}\\
G_{W}[w, \dot{w}]= \int_\Sigma \! \left[ \frac{\dot{w}^{\FO }}{N}\mathcal{P} (x)+ w \left( \mathcal{W}+N\Pi_{N}+\pounds_{\vec{N}}\frac{\mathcal{P}}{N}\right)\right].\label{eq:g3}
\end{eqnarray}
Note that appart from the terms involving $\mathcal{P}$, Eqs. \eqref{eq:g1} and \eqref{eq:g2} assume the same structure as in Einstein gravity \cite{Castellani:1981us}. Of course the generator in \eqref{eq:g3} describing Weyl symmetry is absent in Einstein gravity.

The relation to the covariant formulation of infinitesimal diffeomorphisms generated by a vector field $\xi^a$ on $\mathcal{M}$ is simply the ADM decomposition of $\xi$
\begin{equation}
\xi^a=\epsilon_\perp n^a+\epsilon^a, \quad\mathrm{with}\quad \epsilon^a=h^a_b\xi^b\quad \mathrm{and}\quad \epsilon_\perp=-n_a\xi^a.
\end{equation}
In a coordinate system spanned by $(t,\vec{x})$ one can directly show that the Castellani generators we found indeed generate diffeomorphisms and Weyl rescalings of the metric by a straightforward calculation.
With $g_{tt}=-N^2+N^iN^jh_{ij}$, $g_{tj}=N^ih_{ij}$ and $g_{ij}=h_{ij}$ together with $\epsilon_\perp=N\xi^t$ and $\epsilon^i=\xi^i+N^i\xi^t$ one obtains
\begin{eqnarray}
\left\{g_{\mu\nu},G_W[w]\right\}=2wg_{\mu\nu},\\
\left\{g_{\mu\nu},G_\perp[\epsilon_\perp]+G_D[\vec{\epsilon}]\right\}=\pounds_\xi g_{\mu\nu}.
\end{eqnarray}
The generators we obtained differ from the generators on the full phase space obtained in \cite{Kluson:2013hza}. Most notably our generator of Weyl transformations leaves the shift vector field unchanged and includes the action of the primary first class constraint $\mathcal{P}$ necessary to transform $K_{ab}$ and $\Pi_h^{ab}$ correctly.

Another nice property of the Castellani gauge generators is the modification of the well-known surface deformation (SD) algebra which can be read off the Poisson brackets among the constraints $H_\perp$ and $V$ in Eqs. \eqref{eq:DiracAlg}. The SD algebra is
\begin{eqnarray}
\left[\xi,\chi\right]_{\mathrm{SD}}^\perp&=&\pounds_\epsilon\eta_\perp,\nonumber\\
\left[\xi,\chi\right]_{\mathrm{SD}}^a&=&h^{ab}\left(\epsilon_\perp D_b\eta_\perp- \eta_\perp D_b\epsilon_\perp\right)+\pounds_\epsilon\eta^a,
\end{eqnarray}
where $\xi^a=\epsilon_\perp n^a+\epsilon^a$ and $\chi^a=\xi^a=\eta_\perp n^a+\eta^a$.

The modification of the SD algebra is due to taking the action of the PFCs into account. By calculating the Poisson brackets of $G[\xi]:=G_\perp[\epsilon_\perp]+G_D[\vec{\epsilon}]$ with $G[\chi]:=G_\perp[\eta_\perp]+G_D[\vec{\eta}]$ one finds
\begin{equation}
\left\{G[\xi],G[\chi]\right\}=G[[\xi,\chi]]+PFC,\label{eq:CastellaniAlg}
\end{equation}
with
\begin{eqnarray}
\left[\xi,\chi\right]^\perp&=&n_a\ ^{\ms{(4)}}\pounds_\xi\chi^a,\nonumber\\
\left[\xi,\chi\right]^a&=&h^a_b\ ^{\ms{(4)}}\pounds_\xi\chi^b.
\end{eqnarray}
In order to treat the variations of $\dot{\epsilon}_\perp$ and $\dot{\epsilon}^a$ correctly, we set $\dot{N}=\lambda_N$ and $\dot{N}^a=\lambda_{\vec{N}}^a$.

\section{Canonical charges}\label{section3}
\subsection{Boundary conditions}\label{secBCs1}
We assume that the manifold $\mathcal{M}$ has a timelike boundary with topology $\mathbb{R}\times\partial\Sigma$, with $\partial\Sigma$ being compact. We assume that near the boundary the metric can be written as 
\begin{equation}
g_{ab}=\Omega^2\overline{g}_{ab},\label{eq:BC1}
\end{equation}
with finite $\overline{g}_{ab}$ and arbitrary $\Omega$.

The ADM variables near the boundary $\partial \Sigma$ can then be written as
\begin{equation}
h_{ab}=\Omega^2\overline{h}_{ab},\quad N=\Omega \overline{N}, \quad N^a=\overline{N}^a\!.\label{eq:WeylMetric}
\end{equation}

By evaluating the evolution equations for $h_{ab}$, $K_{ab}$ as well as $\Pi_K^{ab}$ and imposing the constraint $\mathcal{P}$ we find the following (on shell) relations (see also Appendix \ref{app2}) for $K_{ab}$, $\Pi_K^{ab}$ and $\Pi_h^{ab}$:
\begin{eqnarray}\label{eq:WeylRest}
K_{ab}&=&\Omega\left[\frac{\overline{h}_{ab}}{\overline{N}}\left(\partial_t-\pounds_{\vec{N}}\right)\ln\Omega+\overline{K}_{ab}\right]\!,\\
\Pi_K^{ab}&=&\Omega^{-1}\overline{\Pi}_K^{ab},\nonumber\\
\Pi_h^{ab}&=&\Omega^{-2}\left[-\frac{1}{\overline{N}}\left(\partial_t-\pounds_{\vec{N}}\right)\ln\Omega\overline{\Pi}_K^{ab}+\overline{\Pi}_h^{ab}\right]\!.
\end{eqnarray}
Therefore, we allow the variations of the canonical variables at the boundary to be
\begin{eqnarray}
\delta h_{ab}&=&\Omega^2\left(2\delta\ln\Omega\overline{h}_{ab}+\delta\overline{h}_{ab}\right),\\
\delta K_{ab}&=&\Omega\delta\ln\Omega\left[\frac{\overline{h}_{ab}}{\overline{N}}\left(\partial_t-\pounds_{\vec{N}}\right)\ln\Omega+\overline{K}_{ab}\right]+\nonumber\\
&\quad &+\Omega\frac{\overline{h}_{ab}}{\overline{N}}\left(\partial_t-\pounds_{\vec{N}}\right)\delta\ln\Omega+\nonumber\\
&\quad &+\Omega\left\{\frac{\delta\overline{h}_{ab}}{\overline{N}}\left(\partial_t-\pounds_{\vec{N}}\right)\ln\Omega+\delta\overline{K}_{ab}+\right.\nonumber\\
&\quad &\left.+\frac{\overline{h}_{ab}}{\overline{N}}\left[-\left(\partial_t-\pounds_{\vec{N}}\right)\ln\Omega\frac{\delta\overline{N}}{\overline{N}^2}-\delta N^a\overline{D}_a\ln\Omega\right]\right\},\\
\delta \Pi_h^{ab}&=&-\frac{2\delta\ln\Omega}{\Omega^2}\left[-\frac{1}{\overline{N}}\left(\partial_t-\pounds_{\vec{N}}\right)\ln\Omega\overline{\Pi}_K^{ab}+\overline{\Pi}_h^{ab}\right]+\nonumber\\
&\quad &-\frac{1}{\Omega^2\overline{N}}\left(\partial_t-\pounds_{\vec{N}}\right)\delta\ln\Omega\overline{\Pi}_K^{ab}+\nonumber\\
&\quad &+\Omega^{-2}\left\{-\frac{1}{\overline{N}}\left(\partial_t-\pounds_{\vec{N}}\right)\ln\Omega\delta\overline{\Pi}_K^{ab}+\delta\overline{\Pi}_h^{ab}\right.+\nonumber\\
&\quad &\left.+\frac{\overline{\Pi}_h^{ab}}{\overline{N}}\left[\left(\partial_t-\pounds_{\vec{N}}\right)\ln\Omega\frac{\delta\overline{N}}{\overline{N}^2}+\delta N^a\overline{D}_a\ln\Omega\right]\right\},\\
\delta \Pi_k^{ab}&=&\Omega^{-1}\left(-\delta\ln\Omega\overline{\Pi}_K^{ab}+\delta\overline{\Pi}_K^{ab}\right),\label{eq:WeylVari}
\end{eqnarray}
where both $\delta$ and $\ln$ only act on the first term on the right.
For the rescaled variables, i.e. the objects denoted with an overline we demand that
\begin{eqnarray}
\delta \overline{h}_{ab}\vert_{\partial\Sigma}=D_c\delta \overline{h}_{ab}\vert_{\partial\Sigma}=0,\label{eq:WeylMetricVari1}\\
\delta \overline{N}\vert_{\partial\Sigma}=D_c\delta \overline{N}\vert_{\partial\Sigma}=0,\label{eq:WeylMetricVari2}\\
\delta N^a\vert_{\partial\Sigma}=D_c\delta N^a\vert_{\partial\Sigma}=0.\label{eq:WeylMetricVari3}
\end{eqnarray}
Demanding consistency with the equations of motion we find
\begin{eqnarray}
\delta \overline{K}_{ab}\vert_{\partial\Sigma}=0,
\end{eqnarray}
and that $\delta\overline{\Pi}_K^{ab}\vert_{\partial\Sigma}$ as well as $\delta\overline{\Pi}_h^{ab}\vert_{\partial\Sigma}$ are arbitrary but finite.
The gauge transformations that preserve the above set of boundary conditions are the bulk diffeomorphisms $\xi^a$ that satisfy
\begin{equation}
\pounds_\xi\overline{g}_{ab}=2\lambda\overline{g}_{ab},\label{eq:BC2}
\end{equation}
close to the boundary with regular $\xi^a\vert_{\partial\Sigma}$. Additionally we also allow for arbitrary Weyl rescalings
\begin{equation}
\delta_wg_{ab}=2w g_{ab}.\label{eq:BC3}
\end{equation}

\subsection{Functionally differentiable generators and charges}
In order to obtain a well defined action of the canonical gauge generators -- given a boundary value problem such as the one specified above -- one usually has to add to the gauge generators, $G$, suitable boundary terms, $Q$, such that they become functionally differentiable. By that one means that the total variation of the improved generator denoted by $\Gamma$ is given by volume integrals only upon imposing the boundary conditions. Let us start with the generator of Weyl transformations, which we can immediatly improve by partially integrating the last term in \eqref{eq:g3}
\begin{equation}
\Gamma_W[w]= \int_\Sigma \! \left[ \frac{\dot{w}^{\FO }}{N}\mathcal{P}+ w \left( \mathcal{W}+N\Pi_{N}\right)-\frac{\mathcal{P}}{N}\pounds_{\vec{N}}w\right]\label{eq:imprGW}.
\end{equation}
The boundary term appearing in the integration by parts is proportional to $\mathcal{P}$ and thus vanishes on shell. Therefore there exists no canonical charge associated with Weyl symmetry.

Let us continue with the generator of spatial diffeomorphisms. By adding boundary terms to \eqref{eq:g2} that are proportional to $\Pi_N$ and $\Pi^{\vec{N}}_a$ which vanish on shell as well as the boundary term
\begin{equation}
Q_D[\epsilon]=2\int_{\partial\Sigma}\ast\epsilon^c\left(\Pi_h^{ab}h_{bc}+\Pi_K^{ab}K_{bc}\right)\label{eq:diffcharge}
\end{equation}
one obtains the generator
\begin{equation}
\Gamma_D[\epsilon]=\int_{\Sigma}\left(\dot{\epsilon}^{a} {\Pi_{\vec{N}}}_{a}+\Pi_K^{ab}\pounds_\epsilon K_{ab}+\Pi_h^{ab}\pounds_\epsilon h_{ab}+ \Pi_{N}\pounds_\epsilon N+{\Pi_{\vec{N}}}_{a}\pounds_\epsilon N^a\right).\label{eq:imprD}
\end{equation}
We will show below that it is already improved. With Eqs. \eqref{eq:WeylMetric}-\eqref{eq:WeylRest} the term in parenthesis of \eqref{eq:diffcharge} is on shell equivalent to
\begin{equation}
\Pi_h^{ab}h_{bc}+\Pi_K^{ab}K_{bc}=\overline{\Pi}_h^{ab}\overline{h}_{bc}+\overline{\Pi}_K^{ab}\overline{K}_{bc},
\end{equation}
which proves that $Q_D[\epsilon]$ is finite.

The boundary integrals obtained from the total variation of the modified generator \eqref{eq:imprD} either vanish due to the constraints or due to the boundary conditions:
\begin{eqnarray}
\delta \Gamma_D[\epsilon]&\approx&\int_{\partial\Sigma}\ast \epsilon^a \left(\Pi_K^{bc}\delta K_{bc}+\Pi_h^{bc}\delta h_{bc}\right)+\ast\xi^t\delta N^c\left(\Pi_h^{ab}h_{bc}+\Pi_K^{ab}K_{bc}\right)\approx\nonumber\\
&\approx&\int_{\partial\Sigma}\ast \epsilon^a \left(\overline{\Pi}_K^{bc}\delta \overline{K}_{bc}+\overline{\Pi}_h^{bc}\delta \overline{h}_{bc}\right)+\ast\xi^t\delta N^c\left(\overline{\Pi}_h^{ab}\overline{h}_{bc}+\overline{\Pi}_K^{ab}\overline{K}_{bc}\right)=0.
\end{eqnarray}
Therefore we conclude that the improved generator of spatial diffeomorphisms is given by \eqref{eq:imprD} and the associated charge is given by \eqref{eq:diffcharge}.

Finally we improve the generator of transversal diffeomorphisms. Note that the charge we added in the case of $G_D[\epsilon]$ is exactly the same boundary intgeral as in \eqref{Lagrangean3} with the shift $N^a$ replaced by $\epsilon^a$. Likewise we might as a first guess add
\begin{equation}
Q_\perp[\epsilon_\perp]=\int_{\partial\Sigma}\ast\left(\Pi_K^{ab}D_b\epsilon_\perp-D_b\Pi_K^{ab}\epsilon_\perp\right).\label{eq:perpcharge}
\end{equation}
Again using  Eqs. \eqref{eq:WeylMetric}-\eqref{eq:WeylRest} we find that
\begin{eqnarray}
Q_\perp[\epsilon_\perp]\approx\int_{\partial\Sigma}\ast\left(\overline{D}_b\overline{\epsilon}\overline{\Pi}_K^{cb}-\overline{\epsilon}\overline{D}_b\overline{\Pi}_K^{cb}+\overline{\epsilon}\overline{D}^c\ln\Omega\overline{\mathcal{P}}\right),.\label{eq:finiteQperp}
\end{eqnarray}
Thus $Q_\perp[\epsilon_\perp]$ is also proven to be finite on the constraint surface. It remains to be shown that $\Gamma_\perp[\epsilon_\perp]=G_\perp[\epsilon_\perp]+Q_\perp[\epsilon_\perp]$ is the desired improved gauge generator. Note that the only potentially dangerous contributions arise from the total variation of $H_\perp[\epsilon_\perp]$ which is described in detail in Eqs. \eqref{dKH0bound}--\eqref{dhH0bound} of Appendix \ref{app}. In these we plug in Eqs. \eqref{eq:WeylMetric}-\eqref{eq:WeylVari} and find the following non-vanishing contribution
\begin{equation}
\int_{\partial\Sigma}\ast\left[-\overline{\epsilon}\overline{D}^c\ln\Omega\left(\overline{\Pi}_K^{ab}\delta\overline{h}_{ab}+\delta\overline{\Pi}_K^{ab}\overline{h}_{ab}\right)+\overline{\epsilon}\overline{D}_b\delta\overline{\Pi}_K^{cb}-\overline{D}_b\overline{\epsilon}\delta\overline{\Pi}_K^{cb}\right],
\end{equation}
which is exactly canceled by the variation of \eqref{eq:finiteQperp} on the constraint surface.

\subsection{Asymptotic symmetry algebra}
Assuming for the moment that $\xi^a$ and $\chi^a$ are pure gauge one can simply take over the result \eqref{eq:CastellaniAlg} for the improved generators
\begin{equation}
\left\{\Gamma[\xi],\Gamma[\chi]\right\}=\Gamma[[\xi,\chi]]+PFC.\label{eq:imprCastellaniAlg}
\end{equation}
Furthermore, it was proven in \cite{Brown:1986ed} that this statement remains true also in the case where  $\xi^a$ and $\chi^a$ are boundary condition preserving transformations: the Poisson bracket of functionally diffentiable generators, whose action is compatible with the boundary conditions is necessarily an improved generator. Following \cite{Henneaux:1985tv,Brown:1986nw} one might fix a gauge, which turns the first class constraints to second class constraints. These are then required to vanish strongly and the Poisson brackets are converted to Dirac brackets. Note, that the improved generators evaluated on shell are the charges. Therefore, in terms of Dirac brackets we obtain
\begin{equation}
\left\{Q[\xi],Q[\chi]\right\}^\star=Q[[\xi,\chi]],
\end{equation}
where we assumed that the algebra of the boundary condition preseving gauge transformations does not admit any central extension. Consequently the asymptotic symmetry algebra -- the Dirac algebra of the charges -- is isomorphic to the Lie algebra of the boundary condition preserving transformations.

\subsection{Time conservation of the charges}
In Eq. \eqref{eq:imprCastellaniAlg} one might set $\xi^a=t^a$. Although $t^a$ is not a boundary condition preserving generator, $\Gamma[t]$ has a well defined functional derivative. Furthermore, with $\dot{N}=\lambda_N$ and $\dot{N}^a=\lambda_{\vec{N}}^a$ we have $\Gamma[t]\equiv H_T$. This shows that $H_T$, which we obtained from the Legendre transformation of the unmodified Weyl squared action, is functionally differentiable. Therefore, the Hamiltonian equations of motion are well defined in the sense that no additional boundary terms are necessary. This concurs with the observation made in \cite{Grumiller:2013mxa} that no boundary terms are needed on the level of the action in order to obtain a well defined variational principle. Let $\Gamma[\chi]$ be an improved generator with $\chi^a$ compatible with the boundary conditions, then
\begin{equation}
\left\{H_T,\Gamma[\chi]\right\}=\Gamma[-\pounds_\chi t^a]+PFC=\Gamma[\dot{\chi}]+PFC.\label{eq:timecharge}
\end{equation}
Again fixing a gauge and employing the Dirac bracket, the left hand side might be read as a time evolution equation for $-Q[\chi]$ on phase space. However, the action of $H_T$ is insensitive to $\chi$. In order to find the total derivative with respect to time of $Q[\chi]$ one needs to add $-Q[\dot{\chi}]$ to Eq. \eqref{eq:timecharge}. Thus we can write
\begin{equation}
\frac{dQ[\chi]}{dt}=Q[\dot{\chi}]-\left\{H_T,Q[\chi]\right\}^\star=0,
\end{equation}
which proves that the charges are preserved in time. 

\section{Anti--de Sitter boundary conditions}\label{AdSBCs}
Let us now present a particular example in which the conditions prescribed in Eqs. \eqref{eq:BC1}, \eqref{eq:BC2}, and \eqref{eq:BC3} are met. Assume a function $\rho$ that allows to foliate space-time close to the boundary with timelike hypersurfaces defined by $\rho=\mathrm{const.}$ with $\rho=0$ being the location of the boundary. The boundary conditions we impose on the metric are
\begin{equation}
ds^2=\frac{\E^{2\omega}\ell^2}{\rho^2}\left(d\rho^2+\gamma_{ij}dx^idx^j\right),\label{eq:BC4}
\end{equation}
with $\omega$ arbitrary and we assume the same Fefferman--Graham expansion for $\gamma_{ij}$ as in \eqref{eq:FGexpansion} where we fix $\gamma^{\LO}_{ij}$ and $\gamma^{\FO}_{ij}$.

For the boundary condition preserving transformation we consider infinitesimal diffeomorphisms $\pounds_\xi$ and Weyl rescalings $\delta_w$ that preserve \eqref{eq:BC4} to leading as well as next to leading order in $\rho$ up to arbitrary rescalings of $\exp(2\omega)$. Whenever $\pounds_\xi$ or $\delta_w$ act on the Weyl prefactor this requirement is satisfied. Thus, we are left with demanding
\begin{equation}
\pounds_\xi \frac{\ell^2}{\rho^2}\overline{g}_{\mu\nu}=2\lambda\frac{\ell^2}{\rho^2}\overline{g}_{\mu\nu}.
\end{equation}
Assuming an asymptotic expansion
\begin{eqnarray}
\xi^\rho=\xi^\rho_{\LO}+\rho\xi^\rho_{\FO}+\mathcal{O}(\rho^2),\nonumber\\
\xi^i=\xi^i_{\LO}+\rho \xi^i_{\FO}+\mathcal{O}(\rho^2),\nonumber\\
\lambda=\lambda_{\LO}+\rho\lambda_{\FO}+\mathcal{O}(\rho^2),
\end{eqnarray}
one immediatly finds $\xi^\rho_{\LO}=0$, $ \xi^i_{\FO}=0$, $\lambda_{\LO}=0$ and
\begin{equation}
\!{}^{\ms{(2+1)}}\!\pounds_{\xi^k_{\LO}}\gamma^{\LO}_{ij}=2\xi^\rho_{\FO}\gamma^{\LO}_{ij},\label{eq:LOCKV}
\end{equation}
at lowest order. Therefore, $\xi^\rho_{\FO}=1/3 \mathcal{D}_i\xi^i_{\LO}$, where $ \mathcal{D}$ denotes the Levi--Cevita connection associated with $\gamma^{\LO}_{ij}$. The next to leading order equation restricts the possible boundary condition preserving transformations further and reads
\begin{equation}
\pounds_{\xi^k_{\LO}}\gamma_{ij}^{\FO}-\frac{2}{3}\,\gamma_{ij}^{\FO}\mathcal{D}_{k}\epsilon^{k}_{\LO}-4\lambda^{\FO}\gamma_{ij}^{\LO}=0,\label{eq:FOCKV}
\end{equation}
where $\lambda^{\FO}$ is obtained from taking the trace of this equation. For example one might choose a $\gamma^{\LO}_{ij}$ that admits the full 10 dimensional conformal algebra. Then in general the choice of $\gamma^{\FO}_{ij}$ will restrict the asymptotic symmetry algebra to a certain subalgebra.

\subsection{Equivalence of canonical charges and Noether charges}
We will now show that the canonical charges are equivalent to the Noether charges obtained in \cite{Grumiller:2013mxa}. The finite part of the metric near the boundary can be decomposed with respect to the timelike unit normal $n^a$ and the unit normal $u_a=\nabla_a\rho$
\begin{equation}
\overline{g}_{ab}=-n_an_b+u_au_b+\sigma_{ab},
\end{equation}
with $\sigma_{ab}$ being the induced metric on $\partial\Sigma$.

We decompose the Weyl tensor in its electric and magnetic parts with respect to $u_a$
\begin{eqnarray}
\mathcal{E}_{ab}=\perp_uu_cu^dC^c_{\ adb},\quad \quad \mathcal{B}_{abc}=\perp_uu_cC^c_{\ abc}.
\end{eqnarray}
As in the ADM split the fully projected Weyl tensor essentially consists of two parts: The first one, a polynomial of the extrinsic curvature of the timelike hypersurface $\rho=\mathrm{const.}$, is irrelevant due to the Cayley--Hamilton theorem. The second one can be written in terms of the electric part of the Weyl tensor.

The momentum $\Pi_K^{ab}$ close to the boundary is decomposed as
\begin{equation}
\Pi_K^{ab}=2u\wedge\omega_\sigma\left(u^au^b\mathcal{E}_{nn}+2u^{(a}\sigma^{b)c}\mathcal{B}_{ncn}+\sigma^{ac}\sigma^{bd}\mathcal{E}_{cd}-\sigma^{ab}\mathcal{E}_{nn}\right).
\end{equation}
The first term in $Q_\perp$ upon using $\epsilon_\perp=N\xi^t$ and $\xi^t=\xi^t_{\LO}+\mathcal{O}(\rho^2)$ at the boundary becomes
\begin{equation}
\int_{\partial\Sigma}\ast D_b\epsilon_\perp\Pi_K^{ba}=\int_{\partial\Sigma}2\omega_\sigma\epsilon_\perp\left(\mathcal{E}_{nn}u\cdot\partial\ln N-\ ^{(2)}D_b\mathcal{B}^{nbn}\right),
\end{equation}
where $\ ^{(2)}D$ denotes the Levi-Civita connection assiciated with $\sigma_{ab}$ and the second term has been obtained by partial integration. The second part of $Q_\perp$ reads
\begin{eqnarray}
-\int_{\partial\Sigma}\ast\epsilon_\perp D_b\Pi_K^{ba}=-\int_{\partial\Sigma}&2&\omega_\sigma\epsilon_\perp\left(n^cn^du\cdot\partial\mathcal{E}_{cd}+2\mathcal{E}_{nn}\mathcal{K}-4\mathcal{E}^{nb}\mathcal{K}_{bn}-\mathcal{E}^{cd}\mathcal{K}_{cd}\right)\nonumber\\
&-&2\omega_\sigma\ ^{(2)}D_b\mathcal{B}^{nbn}-2\omega_\sigma\epsilon_\perp\mathcal{E}_{nn}u\cdot\partial\ln N
\end{eqnarray}
With the relations
\begin{eqnarray}
\ ^{(2)}D_b\mathcal{B}^{nbn}=-n_an_c\mathcal{D}_b\mathcal{B}^{(ac)b}+\mathcal{B}^{acn}k_{ac},\\
\mathcal{E}_{ab}:=\mathcal{E}^{\SO}_{ab}+\mathcal{O}(\rho),\\
u\cdot\partial\mathcal{E}_{ab}:=\mathcal{E}^{\TO}_{ab}+\mathcal{O}(\rho),\\
\mathcal{B}_{abc}:=-\mathcal{B}^{\FO}_{abc}+\mathcal{O}(\rho),
\end{eqnarray}
where $k_{ab}$ denotes the extrinsic curvature of $\partial\Sigma$ in $\partial \mathcal{M}$, we find
\begin{eqnarray}
Q_\perp=\int_{\partial\Sigma}-2\omega_\sigma n_c\xi^g\left(-n_gn^b\right)&&\left(\mathcal{E}^c_{\TO b}+\mathcal{E}^c_{\SO b}\gamma^{\FO}-2\mathcal{E}^c_{\SO d}\gamma^{\FO d}_b-\mathcal{E}^{\SO}_{bd}\gamma^{cd}_{\FO}+\frac{1}{2}\gamma_{\LO b}^{c}\mathcal{E}^{ad}_{\SO}\gamma^{\FO}_{ad}\right.\nonumber\\
&&\left.+2\omega_\sigma\gamma^{\LO}_{be}\mathcal{D}_d\mathcal{B}^{(ce)d}_{\FO}\right)-4\omega_\sigma \epsilon_\perp\mathcal{B}^{abn}k_{ab}.
\end{eqnarray}
The treatment of the charge $Q_D$ is more involved. First we use the equations of motion to get
\begin{equation}
\Pi_h^{ab}=K\Pi_K^{ab}-2\Pi_K^{e(a}K^{b)}_{\ e}+\frac{1}{2}\Pi_K^{cd}K_{cd}h^{ab}+\omega_h\perp\left(n_en_dn\nabla C^{aebd}-2n_d\nabla_c C^{c(ab)d}\right).
\end{equation}
Decomposing $Q_D$ yields
\begin{eqnarray}
Q_D=\int_{\partial\Sigma}-2\omega_\sigma n_c\xi^g\sigma_g^{\ b}&&\left(\mathcal{E}^c_{\TO b}+\mathcal{E}^c_{\SO b}\gamma^{\FO}-2\mathcal{E}^c_{\SO d}\gamma^{\FO d}_b-\mathcal{E}^{\SO}_{bd}\gamma^{cd}_{\FO}\right.\nonumber\\
&&\left.+2\omega_\sigma\gamma^{\LO}_{be}\mathcal{D}_d\mathcal{B}^{(ce)d}_{\FO}\right)+4\omega_\sigma \epsilon_b\ ^{(2)}D_d\perp_n\mathcal{B}^{(bd)n},
\end{eqnarray}
where the last term is integrated by parts and reads
\begin{equation}
\int_{\partial\Sigma}4\omega_\sigma \epsilon_b\ ^{(2)}D_d\perp_n\mathcal{B}^{(bd)n}=-\int_{\partial\Sigma}4\omega_\sigma \ ^{(2)}D_{(d}\epsilon_{b)}\mathcal{B}^{bdn}.
\end{equation}
Then we decompose a diffeomorphism of the metric $\gamma_{ab}$ with the generator with respect to $n_a$ and $\sigma_{ab}$ and obtain for the spatially projected part
\begin{equation}
\sigma_a^{\ c}\sigma_b^{\ d}\pounds_\xi\gamma_{cd}=2\epsilon_\perp k_{ab}+2\ ^{(2)}D_{(a}\epsilon_{b)}=2\lambda^{\LO}\sigma^{\LO}_{ab}+\mathcal{O}(\rho).
\end{equation}
Using $\sigma^{\LO}_{ab}\mathcal{B}^{abn}_{\FO}=0$ we see that the last term in $Q_D$ cancels the last term in $Q_\perp$. Therefore we obtain for the sum of the charges
\begin{eqnarray}
Q[\xi]=\int_{\partial\Sigma}-2\omega_\sigma n_c\xi^{b}&&\left(\mathcal{E}^c_{\TO b}+\mathcal{E}^c_{\SO b}\gamma^{\FO}-2\mathcal{E}^c_{\SO d}\gamma^{\FO d}_b-\mathcal{E}^{\SO}_{bd}\gamma^{cd}_{\FO}+\frac{1}{2}\gamma_{\LO b}^{c}\mathcal{E}^{ad}_{\SO}\gamma^{\FO}_{ad}\right.\nonumber\\
&&\left.+2\omega_\sigma\gamma^{\LO}_{be}\mathcal{D}_d\mathcal{B}^{(ce)d}_{\FO}\right),
\end{eqnarray}
which agrees with the charges found in \cite{Grumiller:2013mxa} up to an overall factor $4$ which is due to the $1/4$ in the Lagrangean we started with. Also note that in Eq. (22) of Ref. \cite{Grumiller:2013mxa} the second line vanishes identically due to the Cayley--Hamilton theorem.

\subsection{Mannheim-Kazanas-Riegert solution}
We will now apply our results to the Mannheim-Kanzas-Riegert (MKR) metric \cite{Riegert:1984zz,Mannheim:1988dj}, given by
\begin{equation}
ds^2=-k(r)dt^2+\frac{dr^2}{k(r)}+r^2d\Omega_{S^{2}}^2
\end{equation}
where $d\Omega_{S^{2}}^2$ is the line element of the 2-sphere and
\begin{equation}
k(r)=\sqrt{1-12 a M}-\frac{2M}{r}-\Lambda r^2+2 a r,
\end{equation}
which for $a=0$ reduces to Schwarzschild - (A)dS. In the case when $a M\ll1$ this solution agrees with the one obtained in an effective model for gravity at large distances \cite{Grumiller:2010bz}. 
We can rewrite this in terms of the Fefferman-Graham expansion prescribed above with
\begin{eqnarray}
\gamma^{\LO}_{ij}&=&diag(-1,0,\sin{(\theta)}^2) \\
\gamma^{\FO}_{ij}&=&\left(\begin{array}{c c c}0 & 0& 0 \\ 0& -2a &0 \\ 0& 0 & -2a\sin(\theta)^2\end{array}\right)
\end{eqnarray}

The metric at leading order admits the full comformal algebra as a solution to Eq. \eqref{eq:LOCKV}. These are given by
 \begin{eqnarray}
 \xi^{\LO i}_{1}&=&(0,0,1),\\
\xi^{\LO i}_{2}&=&(0, \sin (\phi), \cot (\theta)\cos (\phi)),\\
\xi^{\LO i}_{3}&=&(0, -\cos (\phi), \cot (\theta)\sin (\phi)), \\
\xi^{\LO i}_{4}&=&(1,0,0),\\
\xi^{\LO i}_{5}&=&(0, -\cos (\phi), \cot (\theta)\sin (\phi)),\\
\xi^{\LO i}_{6}&=&(\sin (\theta ) \cos (t) \sin (\phi ),\cos (\theta ) \sin (t) \sin (\phi ),\csc (\theta ) \sin (t) \cos (\phi )),\\
\xi^{\LO i}_{7}&=&(\sin (\theta ) \cos (t) \cos (\phi ),\cos (\theta ) \sin (t) \cos (\phi ),-\csc (\theta ) \sin (t) \sin (\phi )),\\
\xi^{\LO i}_{8}&=&(-\cos (\theta ) \sin (t),\sin (\theta ) \cos (t),0),\\
\xi^{\LO i}_{9}&=&(\sin (\theta ) \sin (t) \sin (\phi ),\cos (\theta ) (-\cos (t)) \sin (\phi ),\csc (\theta ) (-\cos (t)) \cos (\phi )),\\
\xi^{\LO i}_{10}&=&(\sin (\theta ) \sin (t) \cos (\phi ),\cos (\theta ) (-\cos (t)) \cos (\phi ),\csc (\theta ) \cos (t) \sin (\phi )),
 \end{eqnarray}
and they agree with asymptotic isometries in \cite{Henneaux:1985tv} in the limit $1/r=\rho/\ell^2\rightarrow 0$. However, the next to leading order term, $\gamma^{\FO}_{ij}$, preserves only the first four Killing vector fields due to Eq. \eqref{eq:FOCKV} and thus restricts the asymptotic symmetry algebra to the $\mathbb{R}\times \mathcal{SO}(3)$ subalgebra of the full conformal algebra $\mathcal{SO}(2,3)$.
 
The only non-vanishing charge is $Q[\xi^{\ms{(0)}i}_{4}]=Q_\perp[N]$ which agrees with the MKR charge obtained in \cite{Grumiller:2013mxa}
 \begin{equation}
Q[\partial_t]=\frac{M}{\ell^2}-a\frac{(1-\sqrt{1-12aM})}{6}.
\end{equation}

\section{Summary and Outlook}\label{section4}
We perfomed the canonical analysis of four dimensional Weyl squared gravity in section \ref{section1} and constructed all the canonical generators associated with the gauge symmetries of the theory using the Castellani algorithm in section  \ref{section2}. As expected, there are two types of generators: Those that correspond to the diffeomorphisms and those that correspond to Weyl rescalings of the metric. These generators have been improved to have well-defined functional derivatives and the corresponding charges have been constructed, see Eqs. \eqref{eq:diffcharge} and \eqref{eq:perpcharge}. Under the boundary conditions specified in section \ref{section3} they are proven to be integrable, finite and conserved. The Dirac bracket algebra of the charges is isomorphic to the Lie algebra of asymptotic diffeomorphisms.

Intriguingly, we found that there exists no charge associated with Weyl rescalings of the bulk metric. Most notably, this statement is independend of the boundary conditions and implies that in general Weyl rescalings are trivial gauge transformations even at the boundary. It is interesting to compare this with the conformally invariant gravitational Cherns-Simons theory in three dimensions studied in \cite{Afshar:2011qw} since in view of this study our result is unexpected. There it was shown that Weyl charge vanishes as well if the Weyl factor is not allowed to vary freely. However, as soon as the Weyl factor is allowed to fluctuate this is no longer the case. Consequently, the asymptotic symmetry algebra (two copies of the Virasoro algebra) is enhanced by a current $U(1)$-algebra.

Our finding that the Weyl charge vanishes is also in accordance with the analysis done in \cite{Campigotto:2014waa}. There a gauge theoretical formulation of CG was performed and it was shown that the superpotential associated with Weyl rescalings is vanishing, implying that conformal symmetry has a trivial conserved charge. Furthermore, in the Weyl invariant scalar--tensor model analyzed in \cite{Jackiw:2014koa} the Noether current associated with Weyl rescalings is also vanishing, thus showing that Weyl symmetry does not play a dynamical role.

Note that the boundary conditions Eq. \eqref{eq:WeylMetric}, Eqs. \eqref{eq:WeylMetricVari1}-\eqref{eq:WeylMetricVari3} and Eq. \eqref{eq:BC2} that render the charges integrable, finite and conserved are formulated background independently. This allows to use our analysis in various future applications, most notably also for non-AdS holography, provided that the particular boundary conditions are compatible with those described here. In section \ref{AdSBCs} we applied our results to AdS boundary conditions, where we showed that in this case the canonical charges are equivalent to the Noether charges found in \cite{Grumiller:2013mxa}. We also discussed the charge and the asymptotic symmetry algebra of the MKR solution.

Moreover, our analysis shows that it is not necessary to add any counterterms on the level of the action also in the case of our general boundary conditions. The total Hamiltonian obtained solely from the Legendre transform of the Weyl squared action has a well defined functional derivative. Therefore, the Hamitonian equations of motion constitute a well defined initial value problem in the presence of a boundary.

As part of our future research, it would be interesting to further investigate the symmetry algebra of the dual field theory and in particular classify all possible subalgebras of $\mathcal{SO}(2,3)$ subject to the conditions \eqref{eq:LOCKV} and \eqref{eq:FOCKV}. As a next step one would need to find a bulk metric corresponding to the respective choices of $\gamma^{\LO}_{ij}$ and $\gamma^{\FO}_{ij}$.
Furthermore, one could calculate the higher point correlation functions and the $1$-loop partition function in the background given by $\gamma^{\LO}_{ij}$ and $\gamma^{\FO}_{ij}$. Another possible route for future studies is to apply our results to non-AdS holographic setups.
 
\acknowledgments We are grateful to  H.~Afshar, R.~McNees and  P.~Mezgolits for discussions. In particular we want to thank D.~Grumiller for many enlightening discussions and guidance. Some calculations have been performed with the \textit{xAct} \cite{xAct} and \textit{RGTC} \cite{RGTC} packages for \textit{Mathematica}. This work was supported by the START project Y 435-N16 of the Austrian Science Fund (FWF). IL was also supported by the FWF project I 952-N16. MI and FP were also supported by the FWF project P 26328-N27.
\appendix
\section{ADM decomposition of the Weyl tensor}\label{ADMWeyl}
From the ADM decomposition of the metric we derive the Gau\ss\ relation
\begin{equation}
\perp\!{}^{\ms{(4)}}R_{abcd}=-K_{ad}K_{bc}+K_{ac}K_{bd}+R_{abcd},
\end{equation}
the Codazzi relation
\begin{equation}
\perp\! n^d\!{}^{\ms{(4)}}R_{abcd}=D_aK_{bc}-D_bK_{ac},
\end{equation}
and the Ricci relation
\begin{equation}
\perp \!n^bn^d\!{}^{\ms{(4)}}R_{abcd}=K_a^{\ e}K_{ec}-\frac{1}{N}\dot{K}_{ac}+\frac{1}{N}D_aD_bN+\frac{1}{N}\pounds_NK_{ac}.
\end{equation}
From that we derive the relations for the Ricci tensor
\begin{eqnarray}
\perp\!{}^{\ms{(4)}}R_{ab}&=&-2K_{ac}K_b^{\ c}+K_{ab}K+\frac{1}{N}\dot{K}_{ab}-\frac{1}{N}\pounds_NK_{ab}+\frac{1}{N}D_aD_cN+R_{ab},\nonumber\\
\perp\!n^b\!{}^{\ms{(4)}}R_{ab}&=&D_cK_a^{\ c}-D_aK,\nonumber\\
n^an^b\!{}^{\ms{(4)}}R_{ab}&=&K_{ab}K^{ab}-\frac{1}{N}h^{ab}\left(\dot{K}_{ab}-\pounds_NK_{ab}\right).
\end{eqnarray}
Finally we find the curvature scalar to be
\begin{equation}
{}^{\ms{(4)}}R=-3K_{ab}K^{ab}+K^2+\frac{2}{N}h^{ab}\left(\dot{K}_{ab}-\pounds_NK_{ab}\right)+R.
\end{equation}
Now let us turn to the Weyl tensor.

First note that since the Weyl tensor is traceless we deduce
\begin{eqnarray}
h^{bd}\perp\!C_{abcd}&=&\perp\!n^bn^dC_{abcd},\\
h^{bd}\perp\!n^aC_{abcd}&=&0,\\
h^{bd}\perp\!n^an^cC_{abcd}&=&0.
\end{eqnarray}
Using the symmetries of the Weyl tensor the first relation together with the third one above allows us to write the trace part of the spatial projection of the Weyl tensor, $\perp\!C_{abcd}$, as
\begin{equation}
h_{bd}\perp n^en^fC_{aecf}+h_{bc}\perp n^en^fC_{aefd}+h_{ad}\perp n^en^fC_{ebcf}+h_{ac}\perp n^en^fC_{ebfd}.\label{Weyltrace}
\end{equation}
One can then work on the traceless part denoted by $K_{abcd}$ separately. The advantage of this separation is that one only needs to impose the Gauss relation together with the tracelessness condition. Therefore, only the extrinsic curvature can be expected to be involved in the final result because the traceless part of the induced Riemann tensor is the induced Weyl tensor, which vanishes identically. Thus one finds
\begin{eqnarray}
K_{abcd}&=&\frac{1}{2}K_{ac}K_{bd}+h_{ac}\left(K_{be}K_d^{\ e}-K_{bd}K\right)-\frac{1}{4}h_{ac}h_{bd}\left(K_{ef}K^{ef}+K^2\right)+\nonumber\\&\ &+(a\leftrightarrow b,c\leftrightarrow d)-(a\leftrightarrow b)-(c\leftrightarrow d).
\end{eqnarray}
The other projections of the Weyl tensor give
\begin{eqnarray}
&&\perp n^dC_{abcd}=2\mathcal{S}_{abc}^{def}D_dK_{ef}=:B_{abc},\label{magneticWeyl}\\
&&\perp n^an^cC_{abcd}=\frac{1}{2}\mathcal{T}_{bd}^{ef}\left[R_{ef}+K_{ef}K-\frac{1}{N}\left(\dot{K}_{ef}-\pounds_NK_{ef}-D_eD_fN\right)\right],
\end{eqnarray}
with the projectors
\begin{eqnarray}
\mathcal{S}_{abc}^{def}&=&h_{a}^{\ [d}h_{b}^{\ e]}h_{c}^{\ f}-h_{a}^{\ [d}h_{bc}h^{e]f}\\
\mathcal{T}_{ab}^{de}&=&2h_{(a}^{\ d}h_{b)}^{\ e}-\frac{1}{3}h_{ab}h^{de}.\label{tracelessproj}
\end{eqnarray}
The decomposition of the Weyl tensor is then split into contributions of the following type:
\begin{eqnarray}
1&\times&\quad \perp\!C_{abcd},\nonumber\\
4&\times&\quad\ n_bn_d\perp\!n^en^fC_{aecf},\nonumber\\
4&\times&\quad -n_a\perp\!n^eC_{ebcd}.
\end{eqnarray}
Using that the Weyl tensor is traceless and exploiting its symmetries we find that in the expansion of $C_{abcd}C^{abcd}$ each of the terms above only contributes when contracted with itself. Thus
\begin{equation}
C_{abcd}C^{abcd}=\perp\!C_{abcd}\perp\!C^{abcd}-4\perp\!n^eC_{ebcd}\perp\!n_fC^{fbcd}+4\perp\!n^en^fC_{aecf}\perp\!n_gn_hC^{agch}
\end{equation}
Each of the 4 terms in \eqref{Weyltrace} gives
\begin{equation}
h_{bd}\perp\!n^en^fC_{aecf}\perp\!C^{abcd}=\perp\!n^en^fC_{aecf}\perp\!n_gn_hC^{agch}.
\end{equation}
In $K_{abcd}K^{abcd}$ only the term $2K_{abcd}K^{ac}K^{bd}$ survives. Due to the Cayley--Hamilton theorem this contribution also vanishes. Note that $-1/3K^a_{\ \ bcd}K^{bd}$ in  matrix form (suppressing indices) reads
\begin{equation}
K^3-K^2\tr K +K \frac{1}{2}\left[\left(\tr K\right)^2-\tr K^2\right]-\Id\frac{1}{6}\left[\left(\tr K\right)^3-3\tr K\tr K^2+2\tr K^3\right],
\end{equation}
which is the characteristic polynomial of $K$ with $K$ as its argument.

\section{Total variation of the momentum and the scalar constraints}\label{app}
The action of the smeared vector constraint $\{\cdot,V[\vec{X}]\}$ when acting on any tensorial density depending on the reduced phase space ($h$ , $K$, $\Pi_h$,$\Pi_K$) only is simply given by
\begin{equation}
\left\{\Phi,V[\vec{\lambda}]\right\}=\pounds_{\vec{\lambda}}\Phi.\label{diffaction}
\end{equation}
The key ingredient for efficiently calculating this bracket uses the fact that a scalar density $\Psi$ can be regarded as a form of maximal degree on a manifold, thus
\begin{equation}
\pounds_{\vec{\lambda}} \Psi=\mathrm{d}(\iota_{\vec{\lambda}}\Psi)+\iota_{\vec{\lambda}}\mathrm{d}\Psi=\mathrm{d}(\iota_{\vec{\lambda}}\Psi).
\end{equation}
Therefore we have
\begin{equation}
\int_\Sigma Y_{a_1\cdots a_n}\pounds_{\vec{\lambda}} \Psi^{a_1\cdots a_n}=-\int_\Sigma \pounds_{\vec{\lambda}} Y_{a_1\cdots a_n}\Psi^{a_1\cdots a_n},
\end{equation}
up to boundary terms. Hence we may treat the Lie derivative analogously to partial integration. For the variations of $V[\vec{\lambda}]$ we obtain
\begin{eqnarray}
\delta_h V[\vec{\lambda}]&=&-\int_\Sigma \pounds_{\vec{\lambda}}\Pi_h^{ab}\delta h_{ab}+\int_{\partial \Sigma}\ast\left(\lambda^c\Pi_h^{ab}-2\Pi_h^{c(a}\lambda^{b)}\right)\delta h_{ab},\label{dhdiffbound}\\
\delta_{\Pi_h}V[\vec{\lambda}]&=&\int_\Sigma\pounds_{\vec{\lambda}}h_{ab}\delta\Pi_h^{ab}-2\int_{\partial \Sigma}\ast\lambda^ch_{bc}\delta\Pi_h^{ab},\label{dPihdiffbound}\\
\delta_K V[\vec{\lambda}]&=&-\int_\Sigma\pounds_{\vec{\lambda}}\Pi_K^{ab}\delta K_{ab}+\int_{\partial \Sigma}\ast\left(\lambda^c\Pi_K^{ab}-2\Pi_K^{ca}\lambda^{b}\right)\delta K_{ab},\label{dKdiffbound}\\
\delta_{\Pi_K}V[\vec{\lambda}]&=&\int_\Sigma\pounds_{\vec{\lambda}}K_{ab}\delta\Pi_K^{ab}-2\int_{\partial \Sigma}\ast\lambda^cK_{bc}\delta\Pi_K^{ab},\label{dPiKdiffbound}
\end{eqnarray}
Using these formulas we find
\begin{eqnarray}
\left\{\Phi,V[\vec{\lambda}]\right\}&=&\int_\Sigma\pounds_{\vec{\lambda}}h_{ab}\frac{\delta}{\delta h_{ab}}\Phi-\pounds_{\vec{\lambda}}\Pi_h^{ab}\frac{\delta}{\delta \Pi_h^{ab}}\Phi\cdots=\nonumber\\
&=&\Phi(h+\delta_\lambda h,\Pi_h+\delta_\lambda\Pi_h,\cdots)-\Phi(h,\Pi_h,\cdots),
\end{eqnarray}
proving Eq. \eqref{diffaction}, which nicely shows that $V[\vec{\lambda}]$ generates spatial diffeomorphisms on the reduced phase space.

The variations of the scalar constraint $H[\lambda]$ read

\begin{eqnarray}
\delta_KH_0[\lambda]&=&\int_\Sigma \lambda\left(\Pi_K^{ab}K+\Pi_K\cdot K h^{ab}+2\Pi_h^{ab}+4\omega_hD_cB^{cab}\right)\delta K_{ab}+\nonumber\\
&\ &D_c\lambda4\omega_hB^{cab}\delta K_{ab}+\nonumber\\
&\ &\int_{\partial\Sigma} \lambda\ast4\omega_hB^{cab}\delta K_{ab},\label{dKH0bound}\\
\delta_{\Pi_h}H_0[\lambda]&=&\int_\Sigma \lambda2K_{ab}\delta \Pi_h^{ab},\\
\delta_{\Pi_K}H_0[\lambda]&=&\int_\Sigma \lambda \left(-\omega_h^{-1}\Pi^K_{ab}+R_{ab}+K_{ab}K\right)\delta\Pi_K^{ab}+\nonumber\\
&\ &+D_aD_b\lambda\delta\Pi_K^{ab}+\nonumber\\
&\ &\int_{\partial\Sigma}\ast\left(\lambda D_b\delta\Pi_K^{ab}-D_b\lambda \delta\Pi_K^{ab}\right),\label{dPiKH0bound}
\end{eqnarray}
and
\begin{eqnarray}
\delta_h H_0[\lambda]&=&\int_\Sigma \lambda\left\{ \omega_h^{-1}\left(\frac{1}{4}\Pi_K\cdot \Pi_Kh^{ab}-\Pi_K^{ac}\Pi_{K\ c}^b\right)-\Pi_K\cdot K K^{ab}+\right.\nonumber\\
&\ &+D_cD^{(b}\Pi_K^{a)c}-\frac{1}{2}h^{ab}D_cD_d\Pi_K^{cd}-\frac{1}{2}D^2\Pi_K^{ab}+\nonumber\\
&\ &+2\omega_h\left[-\frac{1}{4}B\cdot Bh^{ab} +B^{acd}B_{\ cd}^b+\frac{1}{2}B^{cda}B_{cd}^{\ \ b}+\right.\nonumber\\
&\ &\quad\quad\quad  +B^{d(ab)}D_dK-B^{d(ab)}D_cK_d^{\ c}+\nonumber\\
&\ &\quad\left.\left.\quad\quad -D_c\left(B^{d(ab)}K_d^{\ c}+B^{cd(a}K^{b)}_{\ d}+B^{(a\vert dc \vert}K^{b)}_{\ d} \right)\right]\right\}\delta h_{ab}+\nonumber\\
&\ &+D_c\lambda\left[2D_d\Pi_K^{d(a}h^{b)c}+ D^{(b}\Pi_K^{a)c}-\frac{3}{2}D^c\Pi_K^{ab}-D_d\Pi_K^{cd}h^{ab}+\right.\nonumber\\
&\ &\quad \left.\ -2\omega_h\left(B^{d(ab)}K_d^{\ c}+B^{cd(a}K^{b)}_{\ d}+B^{(a\vert dc \vert}K^{b)}_{\ d} \right)\right]\delta h_{ab}\nonumber\\
&\ &+D_cD_d\lambda\left[2\Pi_K^{(d\vert (a}h^{b)\vert d)}-\Pi_K^{ab}h^{cd}-\frac{1}{2}\Pi_K^{cd}h ^{ab}\right]\delta h_{ab}+\nonumber\\
&\ &\int_{\partial\Sigma} \ast\left[2\lambda\omega_h\left(B^{cd(a}K^{b)}_{\ d}+B^{(a\vert dc}K^{b)}_{\ d}+B^{d(ab)}K_d^{\ c}\right)\delta h_{ab}\right.+\nonumber\\
&\ &+\lambda\left(2\delta C^c_{ed}\Pi_K^{ed}-\delta C^e_{de}\Pi_K^{cd}\right)+\nonumber\\
&\ &+\lambda\left(-D^a\Pi_K^{ec}+\frac{1}{2}D^c\Pi_K^{ea}+\frac{1}{2}D_d\Pi_K^{cd}h^{ea}\right)\delta h_{ea}+\nonumber\\
&\ &+ \left.\left(-2D^a\lambda\Pi_K^{ec}+D^c\lambda\Pi_K^{ea}+\frac{1}{2}D_d\lambda\Pi_K^{cd}h^{ea}\right)\delta h_{ea}\right],\label{dhH0bound}
\end{eqnarray}
where $C^a_{bc}$ denotes the difference tensor of two neighboring Levi--Cevita connections.

\section{Useful formulas for calculating the conformal behavior of the canonical variables on shell}\label{app2}
We use the well known properties of geometric objects under conformal rescalings of the metric
\begin{eqnarray}
h_{ab}=\Omega^2\overline{h}_{ab},\quad N=\Omega \overline{N}, \quad N^a=\overline{N}^a
\end{eqnarray}
hence
\begin{equation}
\omega_h=\Omega^3\omega_{\overline{h}}.
\end{equation}
From the equation of motion for $h_{ab}$ we find
\begin{eqnarray}
K_{ab}&=&\frac{1}{2N}\left(\partial_t-\pounds_{\vec{N}}\right)h_{ab}=\nonumber\\
&=&\Omega\left[\frac{\overline{h}_{ab}}{\overline{N}}\left(\partial_t-\pounds_{\vec{N}}\right)\ln\Omega+\overline{K}_{ab}\right]
\end{eqnarray}
The difference tensor and the Ricci tensor transform as
\begin{eqnarray}
C^a_{bc}&=&2\delta^a_{(b}\overline{D}_{c)}\ln\Omega-\overline{h}_{bc}\overline{D}^a\ln\Omega+\overline{C}^a_{bc}\\
R_{ab}&=&-\overline{D}_a\overline{D}_b\ln\Omega-\overline{h}_{ab}\overline{h}^{cd}\overline{D}_c\overline{D}_d\ln\Omega+\nonumber\\
&\ &+\overline{D}_a\ln\Omega\overline{D}_b\ln\Omega-\overline{h}_{ab}\overline{h}^{cd}\overline{D}_c\ln\Omega\overline{D}_d\ln\Omega+\overline{R}_{ab}
\end{eqnarray}
In order to find the scaling behavior of the canonical momenta we also need the relation
\begin{eqnarray}
\frac{1}{N}D_aD_bN&=&\overline{D}_a\overline{D}_b\ln\Omega-\overline{D}_a\ln\Omega\overline{D}_b\ln\Omega+\nonumber\\
&\ &+\overline{h}_{ab}\overline{h}^{cd}\overline{D}_c\ln\Omega\overline{D}_d\ln\Omega+\frac{1}{\overline{N}}\overline{h}_{ab}D^c\ln\Omega\overline{D}_c\overline{N}+\frac{1}{\overline{N}}\overline{D}_a\overline{D}_b\overline{N}.
\end{eqnarray}
From the equation of motion for $K_{ab}$ and demanding $\mathcal{P}=0$ we compute $\Pi_K^{ab}$
\begin{eqnarray}
\Pi_K^{ab}&=&\omega_h\mathcal{T}^{abcd}\left[R_{cd}+K_{cd}K+\frac{1}{N}D_aD_bN-\frac{1}{N}\left(\partial_t-\pounds_{\vec{N}}\right)K_{ab}\right]=\nonumber\\
&=&\Omega^{-1}\overline{\Pi}_K^{ab}.
\end{eqnarray}
This can also be deduced from the projection of the 4d Weyl tensor using $n_a=\Omega \overline{n}_a$ in $\Pi_K^{ac}=\omega_hn^bn^dC^{a\ c}_{\ b\ d}$ and the well known relation $C^a_{\ bcd}=\overline{C}^a_{\ bcd}$.

Finally we need the magnetic part of the 4d Weyl tensor for which we find
\begin{equation}
B_{abc}=\Omega \overline{B}_{abc}.
\end{equation}
The equation of motion for $\Pi_K^{ab}$  can be used to calculate the Weyl rescaling of $\Pi_h^{ab}$ on shell
\begin{eqnarray}
\Pi_h^{ab}&=&-\frac{1}{2N}\left(\partial_t-\pounds_{\vec{N}}\right)\Pi_K^{ab}-\frac{2}{N}D_c\left(N\omega_hB^{c(ab)}\right)-\frac{1}{2}\left(\Pi_K^{ab}K+\Pi_K^{cd}K_{cd}h^{ab}\right)=\nonumber\\
&=&\Omega^{-2}\left[-\frac{1}{\overline{N}}\left(\partial_t-\pounds_{\vec{N}}\right)\ln\Omega\overline{\Pi}_K^{ab}+\overline{\Pi}_h^{ab}\right].
\end{eqnarray}

\bibliographystyle{unsrt}
\bibliography{bibliothek}

\end{document}